\documentclass[superscriptaddress,12pt]{revtex4}

\usepackage{graphicx}
\usepackage{amsmath}
\usepackage{epstopdf}

\newcommand{\BC}{k_{\rm B}}

\begin{document}

\title{Pulling self-interacting polymers in two-dimensions}

\author{J.\ Krawczyk} \email{j.krawczyk@ms.unimelb.edu.au}
\author{I.\ Jensen} \email{i.jensen@ms.unimelb.edu.au}
\author{A.\ L.\ Owczarek} \email{a.owczarek@ms.unimelb.edu.au}
\affiliation{ARC Centre of Excellence for Mathematics and Statistics of Complex Systems,\\ 
Department of Mathematics and Statistics, The University of Melbourne,  Victoria 3010, Australia}
\author{S.\ Kumar} \email{yashankit@yahoo.com}
\affiliation{Department of Physics, Banaras Hindu University, Varanasi 221 005, India}

\begin{abstract}
 We investigate a two-dimensional problem of an isolated
 self-interacting end-grafted polymer, pulled by one end. 
In the thermodynamic limit, we find that the model has only two 
different phases, namely a collapsed phase and a stretched phase. 
We show that the  phase diagram obtained by Kumar {\it at al.\ } 
[Phys. Rev. Lett. {\bf 98}, 128101 (2007)] for small systems, where
differences between various statistical ensembles play an important role, 
differ from the phase diagram obtained here in the thermodynamic limit.
\end{abstract}
\pacs{64.90.+b,36.20.-r,82.35.Jk,87.15.A- }
\maketitle

\section{Introduction \label{sec:Intro}}

The physics of single polymer chains in a poor solvent is still not
very well understood. Away from the $\theta$-temperature, we know that
a polymer will be in either a collapsed or a swollen state
\cite{degennes}. The mean-square radius of gyration $\langle R^2
\rangle_g$ scales with chain length $N$ as $\langle R^2 \rangle_g \sim
const. \times N^{2\nu},$ where $\nu$ is a critical exponent. At low
temperatures, when the polymer is in the collapsed state, $\nu = 1/d$
while at high temperatures an ``extended'' or ``swollen coil'' state
exists where $\nu = 1, 3/4, 0.588\ldots, 1/2$ for
$d=1,2,3,4$ \cite{degennes} respectively. These values are believed to
be exact for $d=1,2$ and, with a logarithmic correction, for $d=4.$
At high temperatures stretching a polymer should produce a state where
$\nu=1$ which we shall refer to as the ``stretched'' state.
Although there are many theoretical
\cite{marenduzzo2003a-a,rosa2003a-a,orlandini2004a-a} as well as
experimental \cite{strick2001a-a} works on pulling of a collapsed
chain, it seems that some issues remain to be fully understood.

\begin{figure}[ht]
\includegraphics[scale=0.7,angle=0]{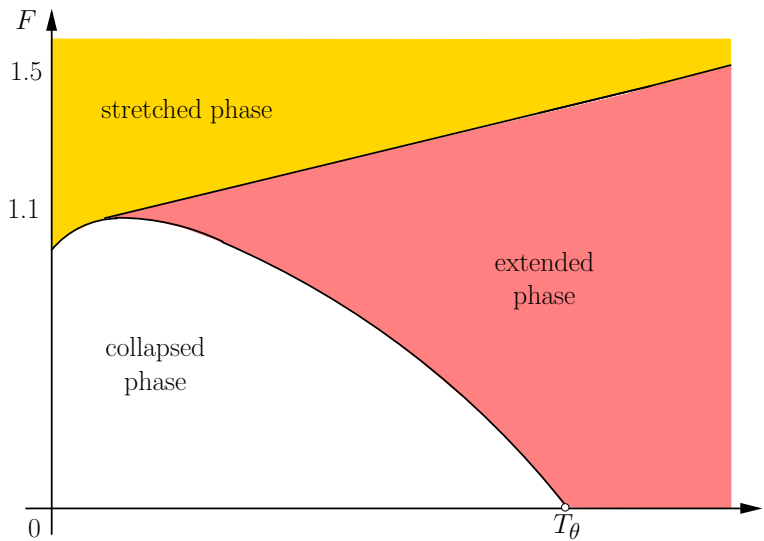}
\caption{\label{fig:pd-prl} 
Schematic phase diagram proposed by Kumar {\it et al.\ }\cite{jensen2007}.
}
\end{figure}

Recently Kumar {\it et al.\ }presented results \cite{jensen2007} 
from exact enumeration of force-induced polymer unfolding in two 
dimension in the context of modeling single molecule experiments. 
For finite systems, they proposed a phase diagram, which 
has three phases namely a collapsed phase, an extended phase and 
a stretched phase (in addition there is a swollen phase which only
occurs at $F=0$ above the $\theta$-temperature). 
They found a transition line between the stretched state and 
the extended phase. The phase diagram proposed is presented in
Fig.~\ref{fig:pd-prl}. The lower phase boundary was obtained 
in both the constant force and constant temperature ensembles, 
and indicates a phase-transition line where polymer goes from the 
collapsed phase to the extended state. However, the upper phase 
boundary was been only in the constant force ensemble and it was
proposed that this represents a transition line where the polymer 
goes from the stretched  state to the extended state.

At this point it is pertinent to mention that very few attempts 
have been made to perform single molecule experiments in the 
constant force ensemble. Only recently did Danilowicz {\it et al.} \cite{danilowicz}
perform stretching experiments on single stranded DNA in the constant 
force ensemble. It was observed that at low force, the extension {\em increases} 
with temperature, while at high force the extension {\em decreases} with 
temperature. Kumar and Mishra \cite{kumar2008} found that this decrease is 
an entropic effect and showed that the upper line is  not a true 
phase transition but a crossover effect.

In this paper we focus our attention on the true nature of the
phase-diagram for the model in the thermodynamic limit.  We present 
some further studies of the
series data trying to gauge the scaling behavior of the model at
different points in the phase-diagram. While somewhat inconclusive our
analysis does indicate that the true phase-diagram (for non-zero
force) has only two distinct phases for non-zero forces and not three
as originally conjectured. The extended phase does not exist for
non-zero forces and the upper phase boundary is a finite-size effect
only present when the model is studied at fixed force with a variable
temperature.

Hence to really delineate the phase diagram we have also performed 
Monte Carlo simulations using the FlatPERM algorithm \cite{prellberg2004}. 
We investigate several hypothetical phase diagrams.  In particular, we 
consider the  possible scenario that the phases
seen are two types of stretched phase; one where the polymer is maximally
stretched in a rod-like conformation and the other where $\nu=1$
though the polymer is not maximally stretched. Using the simulation
results we are able to confidently deduce that there is no evidence of
any additional phase or phase transition. 

We would like to emphasise that the ``phase diagram'' obtained by 
Kumar {\it at al.\ } \cite{jensen2007} for small systems may still be relevant
in the context of experiments on bio-polymers. In real systems of finite size
differences between various statistical ensembles do play an important role
as evidenced not only by this previous study but also by recent experimental 
work \cite{danilowicz}. We thus see our discovery of a discrepancy between the 
finite size ``phase diagram'' and the true infinite size phase diagram as an 
important contribution to a better understanding of the types of finite-size 
effects that may be of importance to the interpretation and understanding of 
experimental results on small systems.

In section~\ref{sec:Model} we define the model.  In
section~\ref{sec:Series} we first briefly review the evidence
presented using series analysis to support the conjectured
phase-diagram \cite{jensen2007,jensen2008} and then present further results from
a more thorough and extensive analysis of the series data casting
doubt on the upper phase boundary of the proposed phase-diagram. In
section~\ref{sec:Simul} we present the conclusive results of the Monte
Carlo simulations which do not support the existence of any additional
phase transitions: we carefully consider various possible scenarios.
Finally, in section~\ref{sec:Summary} we summarize our final
conclusions.

\section{Model  \label{sec:Model}}

\begin{figure}[ht]
\includegraphics[scale=0.7,angle=0]{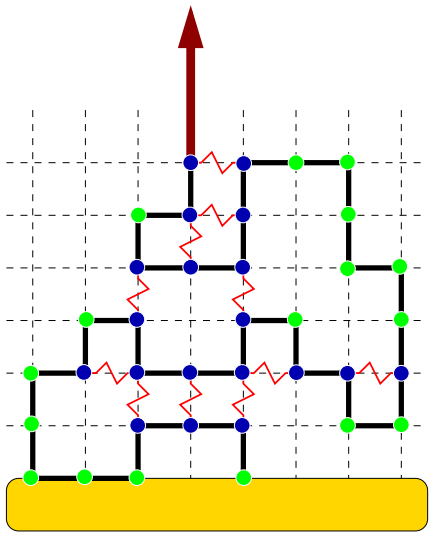} 
\caption{\label{model}
The model of a polymer on the two dimensional square lattice pulled by the 
last monomer. The arrow indicates the direction of the pulling force. 
The dark (red) filled circles on lattice sites denote monomers interacting
via nearest-neighbor interactions.
}
\end{figure}

We model the polymer chains as interacting self-avoiding walks (ISAWs) 
on the square lattice  as shown in Fig.~\ref{model}.
Interactions are introduced between non-bonded nearest neighbor monomers.
In our model one end of the polymer is attached to an  impenetrable neutral 
surface (there are no interactions with this surface) while the polymer 
is being pulled from the other end with a force acting in the direction 
perpendicular to the surface. Note that the ISAW does not extend beyond either 
end-point so the $y$-coordinate $y_j$ of the $j$'th monomer is restricted
by $0=y_0 \leq y_j \leq y_N=h$.

We introduce Boltzmann weights
$\omega=\exp(-\varepsilon/\BC T$) and $u=\exp(F/\BC T)$ conjugate
to the nearest neighbor interactions and force, respectively, where
$\varepsilon$ is the interaction energy,
$\BC$ is Boltzmann's constant, $T$ the temperature and $F$ the
applied force. In the rest of this study we set $\varepsilon=-1$ and $\BC=1$.
We study the finite-length partition functions
\begin{equation}
Z_N(T,F)  =  \sum_{\rm all \ walks}  \omega^m u^h 
      =  \sum_{m,h}  C_{N,m,h}  \omega^m u^h,
\end{equation}
where $C_{N,m,h}$ is the number of ISAWs of length $N$ having $m$ nearest 
neighbor contacts and whose end-point is a distance $h$ from the surface.

\section{Series Analysis \label{sec:Series}}

\subsection{Fluctuation curves and the conjectured phase diagram}

\begin{figure}
\begin{center}
\includegraphics[width=9cm]{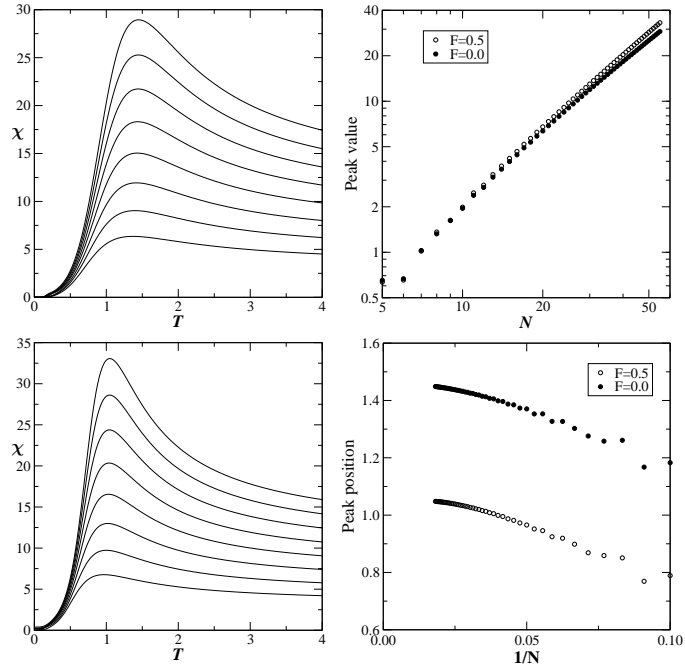}
\end{center}
\caption{\label{fig:FlucF} The fluctuations in the number of contacts
as a function of temperature for fixed force $F=0.0$ (upper left panel)
and $F=0.5$ (lower left panel). Each panel contains curves for ISAWs of
length (from bottom to top) $N=20, 25,\ldots, 55$. In the upper right panel
we show a log-log plot of the growth in the peak value of the fluctuation curve
with chain length $N$. The lower right panel shows the peak position (critical 
temperature value) vs $1/N$. 
}
\end{figure}

To begin let us recall the type of analysis presented by  
Kumar {\it et al.\ }\cite{jensen2007,jensen2008}.
At low temperature and force the polymer
chain is in the collapsed state and as the temperature is increased (at fixed
force) the polymer chain undergoes a phase transition to an extended state.
The value of the transition temperature (for a fixed value of the force) can
be obtained  from the fluctuations in the number of
non-bonded nearest neighbor contacts.
The fluctuations are defined as $\chi=\langle m^2 \rangle -\langle m \rangle^2$,
with the $k$'th moment given by 
$$\langle m^k \rangle = 
\frac{\sum_{m,h} \! m^k C(N,m,h)  \omega^m u^h}{\sum_{m,h} \! C(N,m,h)  \omega^m u^h}.$$
In the panels of Fig.~\ref{fig:FlucF} we show the emergence of peaks in 
the fluctuation curves with increasing $N$ at fixed force $F=0.0$ and $F=0.5$. 
In the top right panel we show the growth in the peak value as $N$ is increased.
Since this is a log-log plot we see that the peak values grows as a power-law with
increasing $N$; this divergence is the hall-mark of a phase transition. In the lower
right panel we have plotted the position of the peak (or transition temperature)
as a functions of $1/N$. Clearly the transition temperature appears to converge to
a finite (non-zero) value but the data exhibits clear curvature which makes an
extrapolation to infinite length difficult.

\begin{figure}
\begin{center}
\includegraphics[width=9cm]{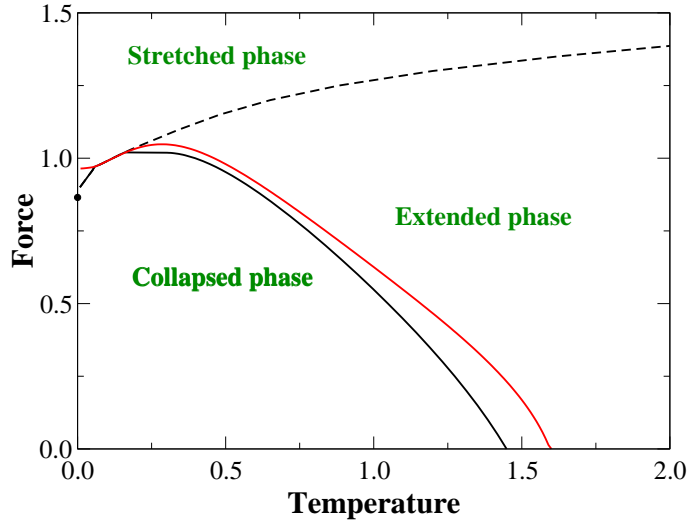}
\end{center}
\caption{\label{fig:phase} The finite-size `phase diagram'  for 
flexible chains as obtained
from the position of the peak in the contact fluctuation curves for $N=55$. The solid
black curve and the dashed curve are obtained by fixing the force and varying 
the temperature.}
\end{figure}

In Fig.~\ref{fig:phase}, we show the force-temperature `phase diagram'
for flexible chains as obtained from the peak positions for the finite chains.
However, true phase diagram should be obtained by extrapolating the data to
the $N\to \infty$ limit). 
In Fig.~\ref{fig:phase} we have shown the transitions as obtained
by fixing the force (black curves).
One of the most notable feature of  the phase-diagram is the
{\em re-entrant} behavior but this has been studied and explained
in previous papers \cite{jensen2007,jensen2008}. The other notable feature is that in the fixed force 
case we see an apparent new transition line 
from the extended state to the fully stretched
state which is solely induced by the applied force (the dashed line
in Fig.~\ref{fig:phase}).

\subsection{Further series analysis results }
 
\begin{figure}
\begin{center}
\includegraphics[width=9cm]{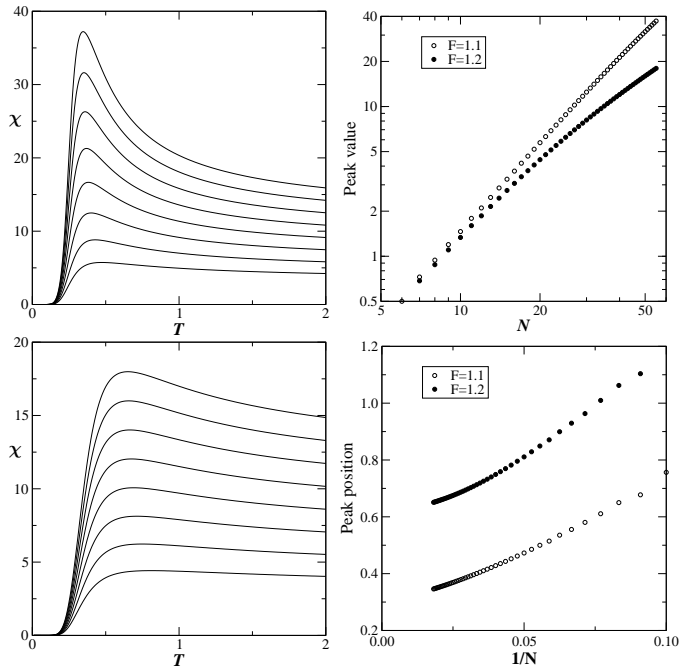}
\end{center}
\caption{\label{fig:FlucHighF} Same as in Fig.~\ref{fig:FlucF} but for
fixed force $F=1.1$ (upper left panel) and $F=1.2$ (lower left panel). 
}
\end{figure}

\begin{figure}
\begin{center}
\includegraphics[width=9cm]{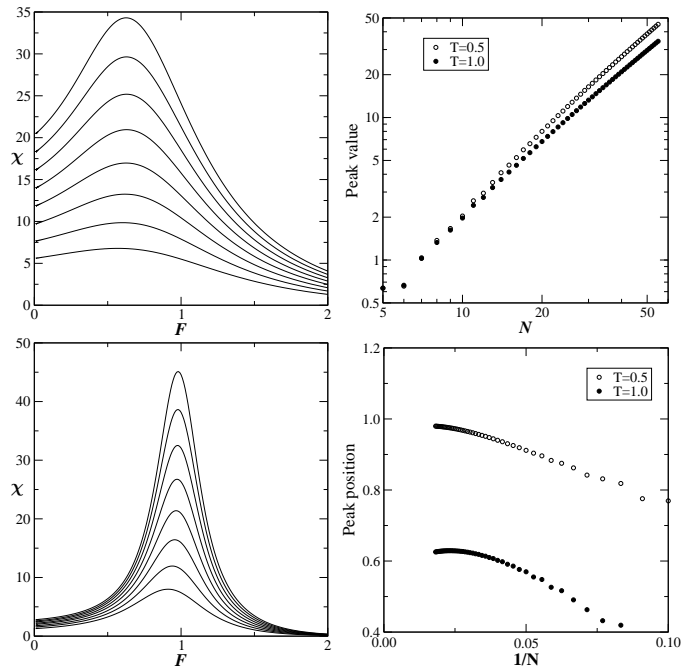}
\end{center}
\caption{\label{fig:FlucT} The fluctuations in the number of contacts
as a function of force for fixed temperature $T=1.0$ (upper left panel)
and $T=0.5$ (lower left panel). Each panel contains curves for ISAWs of
length (from bottom to top) $N=20, 25,\ldots, 55$. In the upper right panel
we show a log-log plot of the growth in the peak value of the fluctuation curve
with chain length $N$. The lower right panel shows the peak position (critical 
force value) vs $1/N$. }
\end{figure}

In Fig.~\ref{fig:FlucHighF} we have plotted
the fluctuation curves for force $F=1.1$ and $F=1.2$. The curves for
$F=1.1$ (including the plot of the peak height) looks very similar to
the plots (see Fig.~\ref{fig:FlucT}) for low values of the force. For
force $F=1.2$ the peak is not very pronounced and we are hesitant to
even call it a peak. Also when we look at the peak height vs. $N$
it appears that the curve has two different behaviors for small
and large $N$, respectively. 
This could be a sign of a cross-over behavior.

One can also study the same transition phenomenon by fixing the temperature
and varying the force. In the panels of Fig.~\ref{fig:FlucT} we show the emergence 
of peaks in  the fluctuation curves with increasing $N$ at fixed temperature
$T=1.0$ and $T=0.5$. Again we observe the power-law divergence of the peak-value.
The only other note-worthy feature is that in the plots of the peak position
(critical force value) we observe not only strong curvature but we actually
see a turning point in the curves as $N$ is increased. This feature would make it
impossible (given the currently available chain lengths) to extrapolate this data.
However,  we {\em do not} observe the upper transition line in this study where
we have fixed the temperature and varied the force. Indeed this is clear
from  Fig.~\ref{fig:FlucT} where at fixed $T=0.5$ and $1.0$ we see only
a single peak (giving us points on the red curve in the `phase diagram'
Fig.~\ref{fig:phase}).

In Fig.~\ref{fig:FlucT} the value of the force extends
up to $F=2.0$ and the upper transition (dashed line in the phase diagram)
should appear (if present) as a second peak in the fluctuation curves
of Fig.~\ref{fig:FlucT}. The absence of any evidence of a second peak
is what leads us conclude that we do not see this second transition 
in the fixed $T$ varying $F$ study.

\begin{figure}
\begin{center}
\includegraphics[width=9cm]{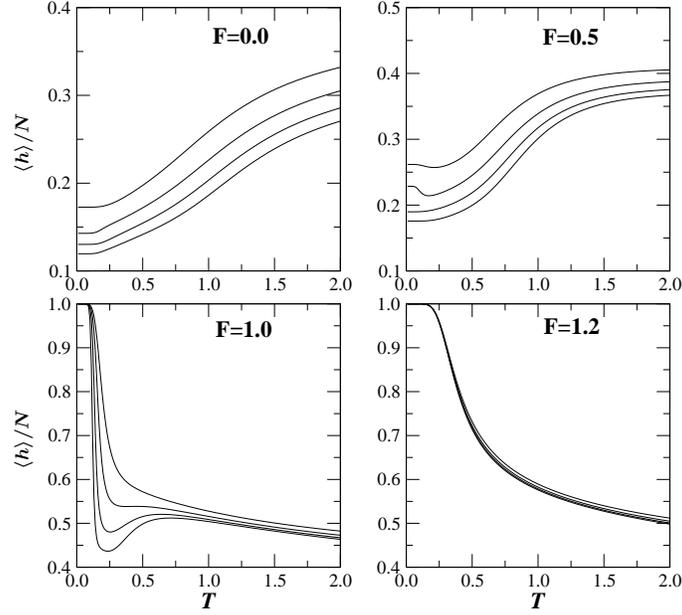}
\end{center}
\caption{\label{fig:AveLen} The average extension per monomer $\langle h \rangle/N$
as a function of temperature $T$ for different values of the force. Each panel
contain four curves for, from top to bottom, $N=25$, 35, 45, and 55.}
\end{figure}

In Fig.~\ref{fig:AveLen} we have plotted the average extension per monomer
$\langle h \rangle/N$ as a function of temperature. In the case of a fixed
force $F=1.2$ (lower right panel) we note that curves for different values of $N$ 
more or less coincide showing that the average extension scales like $N$ for
all temperatures. We contend that this observed behavior shows that the 
upper boundary is a crossover effect supporting the finding reported 
recently by Kumar and Mishra \cite{kumar2008}.
\begin{figure}
\begin{center}
\includegraphics[width=9cm]{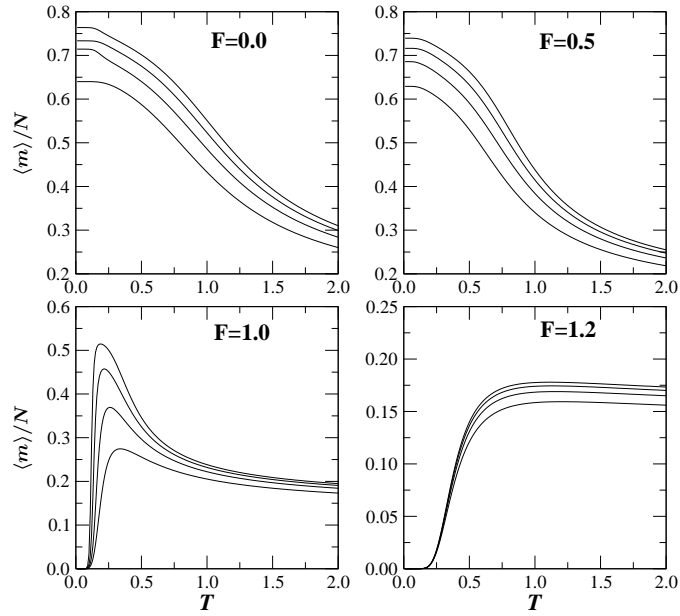}
\end{center}
\caption{\label{fig:AveCont} The average number of contacts per monomer $\langle m \rangle/N$ 
as a function of temperature $T$ for different values of the force. Each panel
contain four curves for, from bottom to top, $N=25$, $35$, $45$, and $55$.}
\end{figure}

In Fig.~\ref{fig:AveCont} we have plotted the average number of contacts per monomer
$\langle m \rangle/N$ as a function of temperature. In the case of a fixed
force $F=1.2$ (lower right panel) we note that curves for different values of $N$ 
more of less coincide showing that the average extension scales like $N$ for
all temperatures.

\section{Simulation results  \label{sec:Simul}}

\begin{figure}[ht!]
\includegraphics[scale=0.6,angle=0]{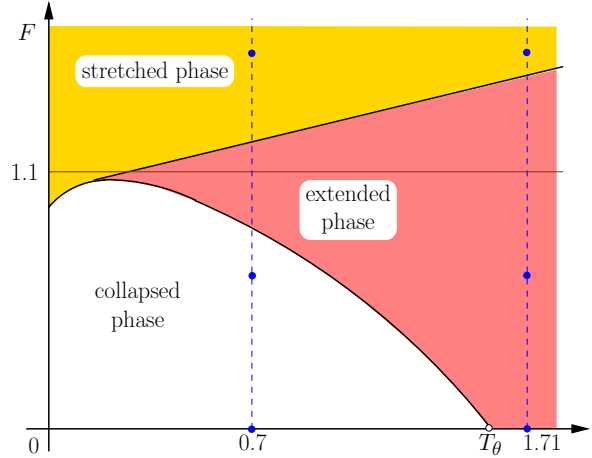}
\caption{\label{pd_conj} 
Schematic phase diagram  conjectured by Kumar {\it et al.\ }\cite{jensen2007, jensen2008}.
The conjectured phase diagram has three different phases: `collapsed', `extended'
and `stretched'. We have performed simulations of the whole phase
space up to length $N_{max}=128$. The dashed lines at fixed temperatures $T=0.7$ and
$T=1.71$ display lines along which simulations were performed for walk
lengths up to $N_{max}=1024$. The six points displayed are those at
which we have focused our attention.}
\end{figure}

Our more detailed analysis of the series indicates that the upper 
phase boundary is not a real phase transition.. To further investigate the model
we have turned to Monte Carlo simulations that allow analysis of
longer polymer chains. We have chosen to use the FlatPERM algorithm
\cite{prellberg2004} to simulate the model.  One advantage of
FlatPERM, as a ``flat histogram'' technique, is the ability to sample
the density of states uniformly with respect to a chosen
parametrisation, so that the whole parameter range is accessible from
one simulation.  This allows us to ``see'' the phase diagram from one
set of results. The cost of this however is that the chain lengths
that can be simulated accurately are still fairly modest. We have
performed ``whole phase space'' simulations up to length $N=128$. On
the other hand by restricting interest to sub-manifolds of the
parameter space longer chains can be analyzed. We have performed
simulations along various lines and at points in the phase diagram
using walks up to length $N=1024$. The schematic phase diagram
conjectured by Kumar {\it et al.\ }\cite{jensen2007,jensen2008} is 
shown in Fig.~\ref{pd_conj} along with special lines considered in our
simulations.

To demonstrate what is estimated in a FlatPERM simulation consider for
a moment a general polymer model with microscopic energies
$-\varepsilon_1$, $-\varepsilon_2$, etc associated with
configurational parameters $m_1$, $m_2$ respectively. Let the density
of states be $C_{N,m_1,m_2,\dots}$.  Then the partition function is
given by
\begin{equation}
Z_N(\beta_1,\beta_2,\dots)=
\sum_{m_1,m_2,\dots} C_{N,m_1,m_2,\dots}\; e^{\beta_1 m_1+\beta_2 m_2+\dots},
\end{equation}
where $\beta_1=\beta\varepsilon_1$, $\beta_2=\beta\varepsilon_2$ etc, and $\beta=\frac{1}{k_B T}$,
with $k_B$  Boltzmann's constant. FlatPERM can estimate
$C_{N,m_1,m_2,\dots}$ or any sum of the $C_{N,m_1,m_2,\dots}$ over any
number of the $m_j$ for a range of lengths $N\leq N_{max}$. If one
finds $C_{N,m_1,m_2,\dots}$ then one can estimate average quantities
over this distribution for any values of $\beta_1,\beta_2,\dots$.
In our model we have $m_1=m$ and $m_2=h$ with
$\beta_1=1/T$ and $\beta_2 =F/T$. We have performed simulations over
the complete space of  the variables $m$ and $h$ for $N\leq
N_{max}=128$. In this way we have estimated the density of states $C_{N,m,h}$.
We performed $10$ different runs with this parametrisation for lengths
up to $N_{max}=128$. We have estimated the average number of contacts per monomer
$\langle m \rangle/N$ and the average extension per step $\langle h
\rangle/N$ and their fluctuations $\sigma^2(m)/N$ and
$\sigma^2(h)/N$. 

Previous work (see \cite{bennett-wood1998} and references therein) has
estimated the $\theta$-point to be around $T=1/0.663 = 1.51$. With
this in mind we have also performed one-parameter simulations with $T$
fixed at $T=0.7$ and at $T=1.71$. The temperatures chosen were to
ensure that one temperature was below and one was above the
$\theta$-temperature as shown in Fig.~\ref{pd_conj}. We also studied the
temperature $T=2.0$ with $F=0$ as a high temperature point.

\begin{figure}[hb]
\includegraphics[scale=0.5,angle=0]{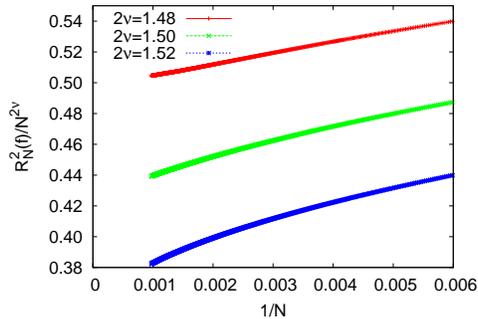}
\caption{\label{ree_ht_fz}
End-to-end distance divided by $N^{2\nu}$ against $1/N$ for point,
$(T,F)= (2.0, 0.0)$. We see $2\nu=1.5(2)$. This point is in the extended phase.
}
\end{figure}

\begin{figure}[hb]
\includegraphics[scale=0.5,angle=0]{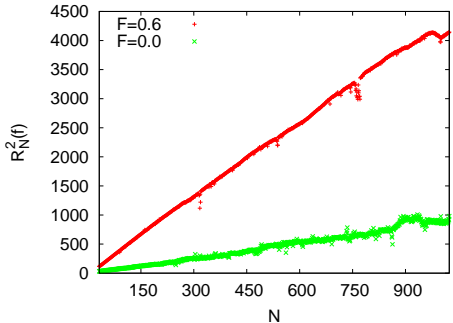}
\caption{\label{ree_coll}
End-to-end distance divided by $N^{2\nu}$ against $N$ for two points,
$(T,F)= (0.7, 0.0)$ and $(T,F)= (0.7, 0.6)$ for lengths up to $N=1024$. A clear linear
dependence is seen implying $2\nu\approx 1$. These points are in the
collapsed phase and this exponent result is consistent with this assumption.
 }
\end{figure}

In order to delineate the possible phases we considered the points
$F=0.0,0.5,1.5$ for $T=0.7$ and $1.71$ and also the point $T=2.0$,
$F=0$. In particular, we analyzed the scaling of the end-to-end
distance $R^2_{N}$ which gives an estimate of the exponent $\nu$. Let
us start with $F=0$. For $T=2.0$ we expect that the polymer is in the
extended phase with $2\nu=1.5$ and in Fig.~\ref{ree_ht_fz} we find
precisely that. For $T=0.7$ we expect the polymer to be in the collapsed phase
with $2\nu=1.0$ and once again our data in Fig.~\ref{ree_coll}
confirms this expectation. Now let us move to $F=0.5$. For the low
temperature $T=0.7$ the series data places this point in the collapsed
phase and the data in Fig.~\ref{ree_coll} bears this out.

\begin{figure}[ht]
\includegraphics[scale=0.5,angle=0]{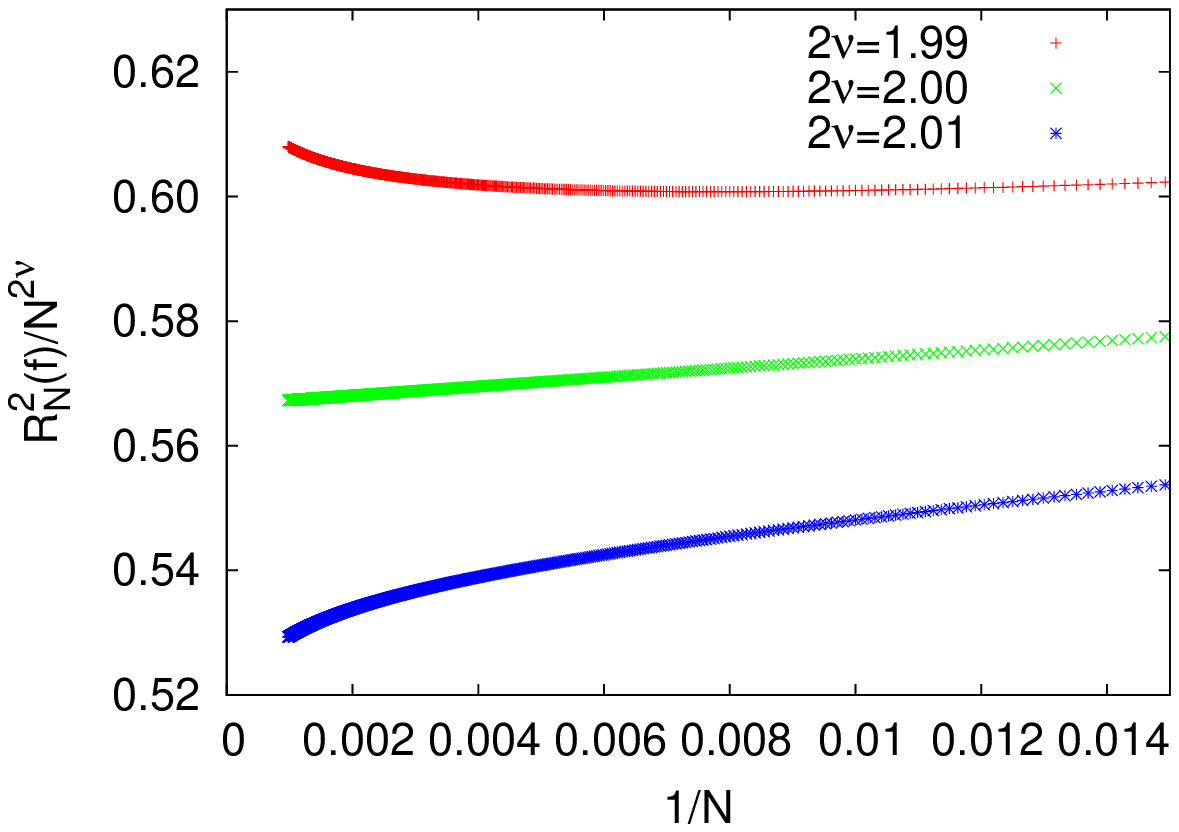}
\includegraphics[scale=0.5,angle=0]{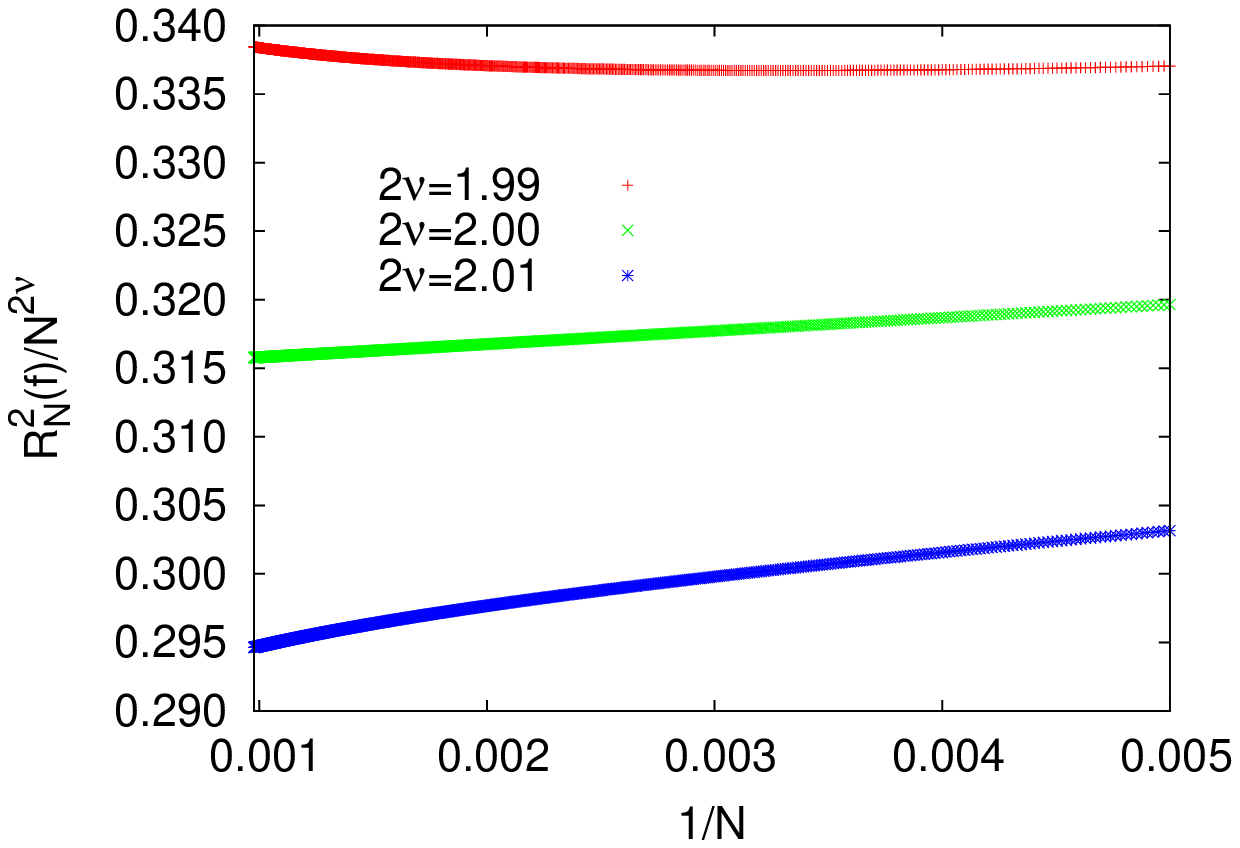}\\
\includegraphics[scale=0.5,angle=0]{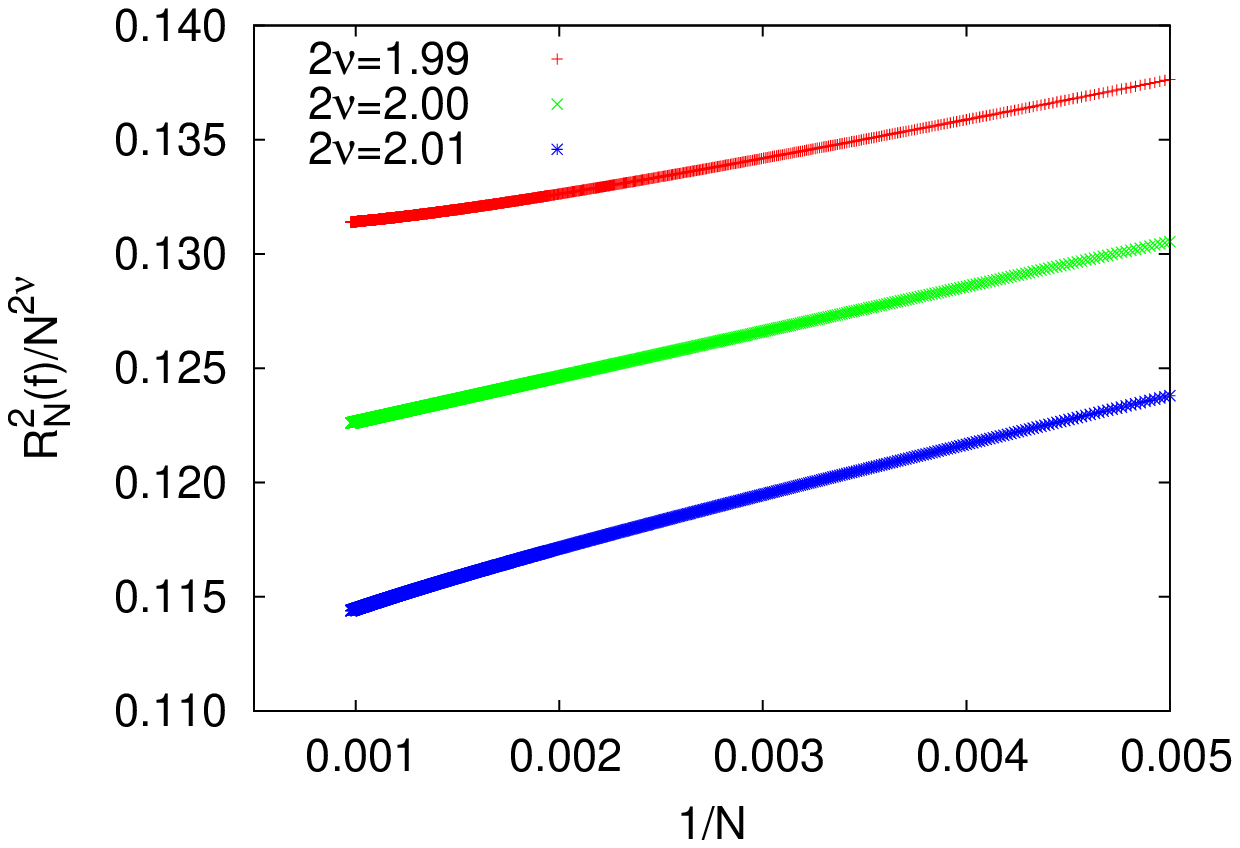}
\caption{\label{ree_stretch}
  End-to-end distance divided by $N^{2\nu}$ against $1/N$ for three
  points, $(T,F)= (0.7, 1.5)$, $(1.71, 1.5)$, and $ (1.71, 0.6)$.  for
  length up to $N=1024$.  We conclude, that all these points belong to
  a stretched phase of same type, where $\nu=1.0$.}
\end{figure}

However, for the point $F=0.5$ with $T=1.71$ the conjectured phase
diagram of Kumar {\it et al.\ }\cite{jensen2007} predicts this point
to be in the extended phase with $2\nu=1.5$ while we find that
$2\nu=2.00(2)$, so this point is in the stretched
phase. For $F=1.5$ and for $T=0.7$ and $T=1.1$ the conjectured phase
diagram predicts a stretched phase with a value of $2\nu=2$ and we confirm
this as seen from Fig.~\ref{ree_stretch}.


\begin{figure}[ht!]
\includegraphics[scale=0.6,angle=0]{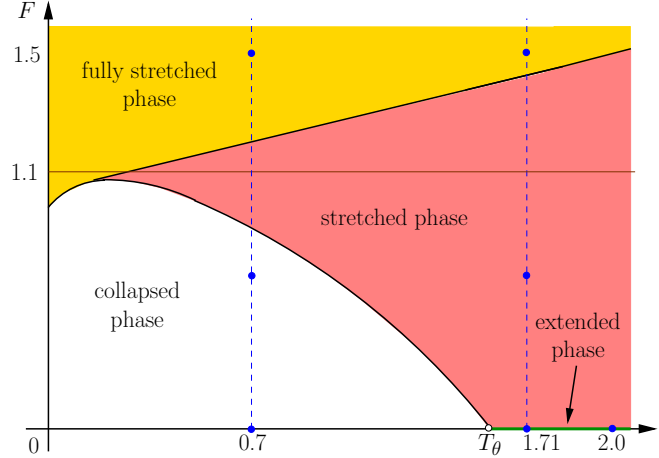}
\caption{\label{pd_inter} 
Here is a second possible phase digram assuming there are two phases
for non-zero forces, both with $\nu=1$. Instead of extended and
stretched we have ``fully stretched'' and stretched. In such a hypothesized ``fully
stretched'' phase the polymer is effectively in a rod-like conformation
where the average height is approximately equal to the length of the polymer.}
\end{figure}

It is clear that the assumption of an extended phase for $F>0$ is
incorrect. However, while mistakenly named perhaps three phases still
exist for $F>0$. Considering the series results the other possibility
is that the extended phase is indeed stretched with $\nu=1$ for $F>0$
and that  the ``stretched'' phase described in Kumar {\it et
  al.\ }\cite{jensen2007,jensen2008} is really a ``fully stretched'' phase where
in addition to $\nu=1$ the average height per step converges to unity:
\begin{equation}
\lim_{N\rightarrow\infty} \frac{\langle h \rangle}{N} =1.
\end{equation}
That is, the configurations of the polymer are essentially rod-like
(with sub-dominant fluctuations). 
For such rod-like configurations one would also expect in this phase that
\begin{equation}
\lim_{N\rightarrow\infty} \frac{\langle m \rangle}{N} =0.
\end{equation}
A revised conjectured phase diagram is drawn in Fig.~\ref{pd_inter},
along with our lines of longer length simulations and the points at
which we have focused our analysis. The series data in 
Fig.~\ref{fig:AveLen} and \ref{fig:AveCont} for the low temperature
regions when $F=1.2$ display behavior resembling that delineated
above for a possible ``fully stretched'' phase.

\begin{figure}[ht]
\includegraphics[scale=0.7,angle=0]{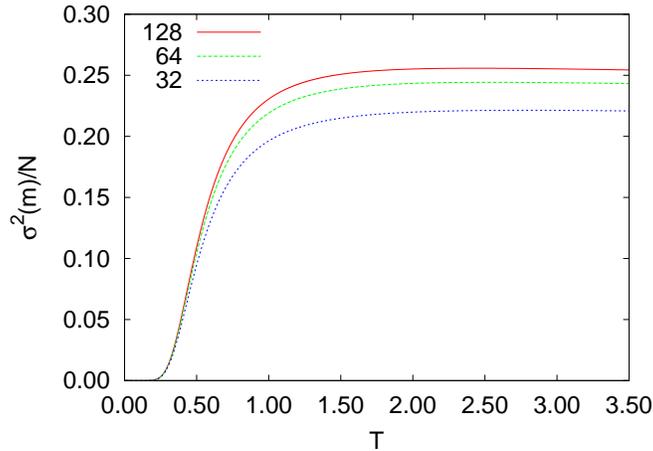}
\caption{\label{f1.4} The fluctuations in the number of contacts $m$ for
  $F=1.4$ fixed plotted against temperature $T$. There is no sign of
  any growing singularities which would indicate a phase transition in
  the thermodynamic limit.
}
\end{figure}

\begin{figure}[ht]
\includegraphics[scale=0.5,angle=0]{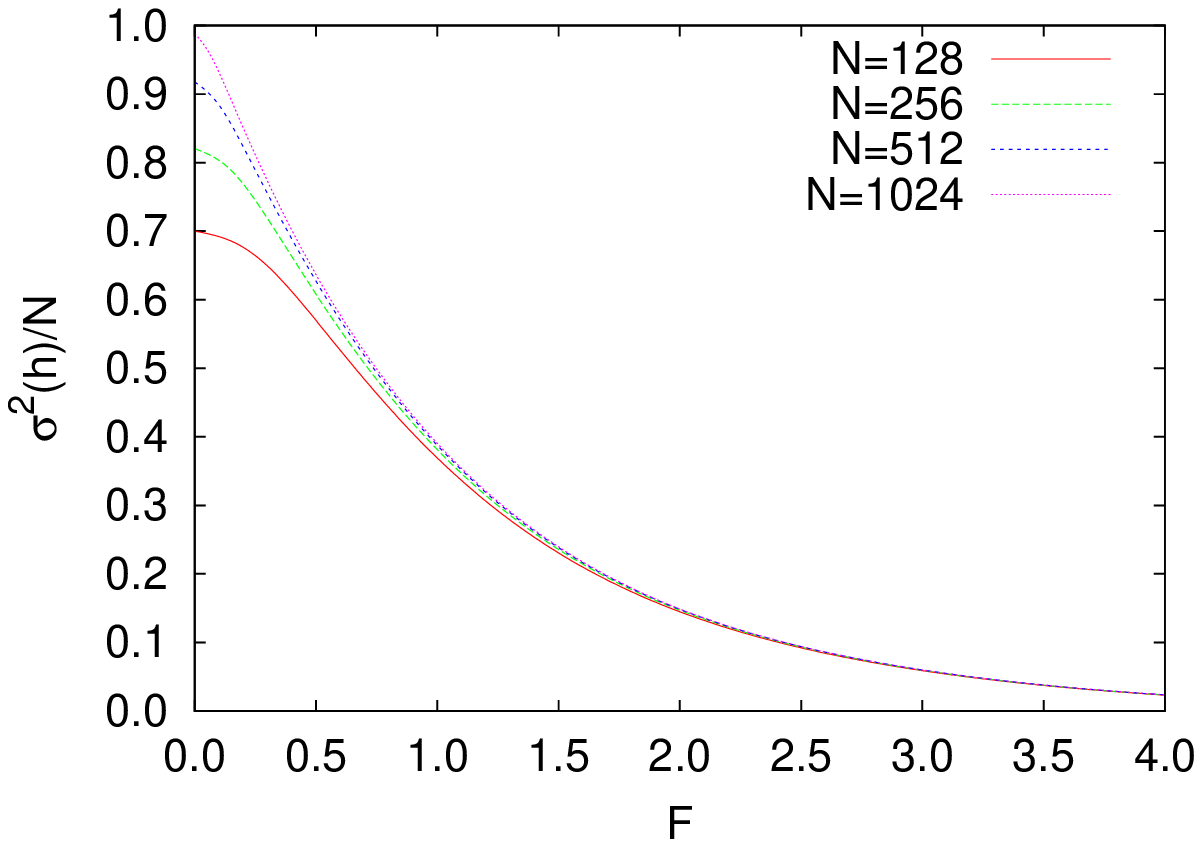}
\includegraphics[scale=0.5,angle=0]{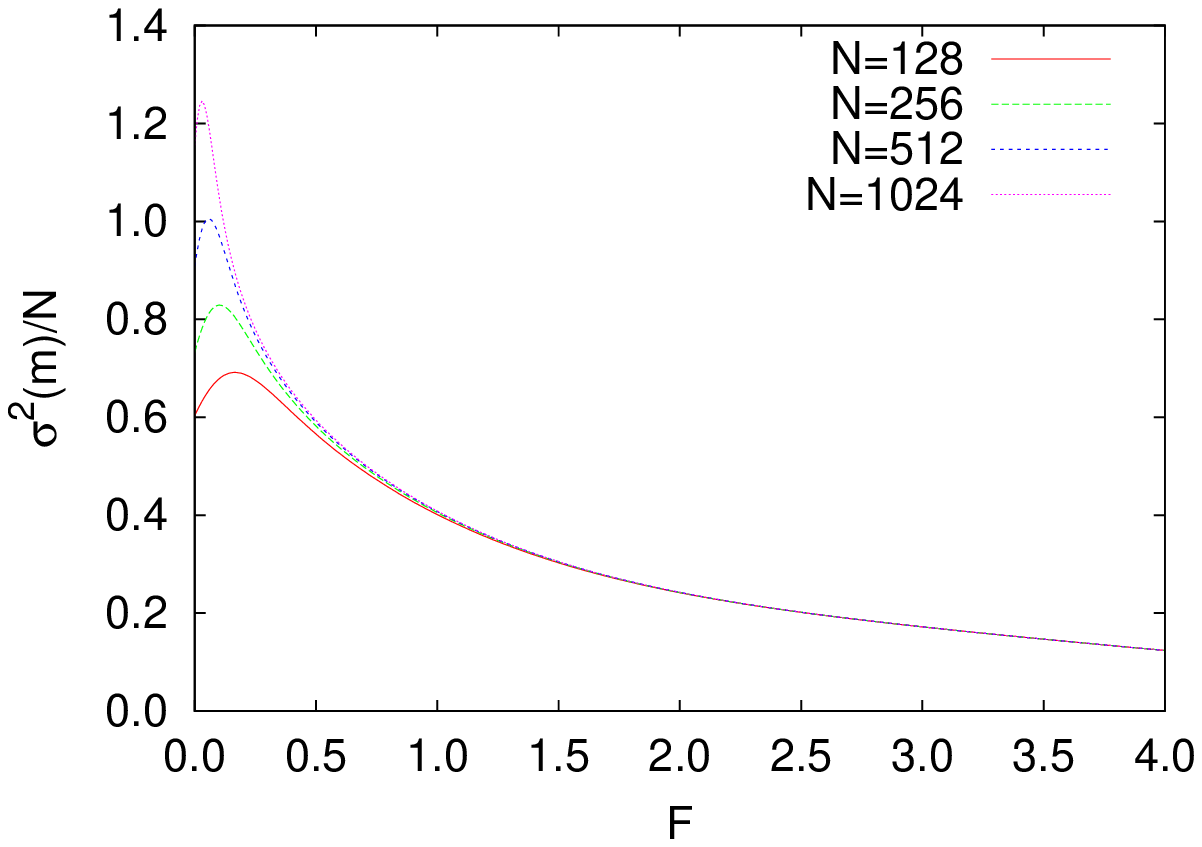}
\caption{\label{flch1.71} Plots of the fluctuations against force $F$
  for the high temperature $T=1.71$.
}
\end{figure}

To search for possible phase transitions we have estimated the
fluctuations in the number of contacts and fluctuations in the height.
For fixed force $F=1.4$ the plot of the fluctuation against temperature
in Fig.~\ref{f1.4} shows no sign of a growing singularity for
lengths up to $N=128$ as seen in the series data for shorter lengths
and smaller forces.
Now we consider the fixed temperature lines at $T=1.71$ and $T=0.7$.  For
$T=1.71$ the only sign of a singularity appears near $F=0$ (see Fig.~\ref{flch1.71}), that is
the  expected sign of the transition from the extended phase at $F=0$ to the
stretched phase at $F>0$.
For $T=0.7$ again there is only a sign of a single phase transition in
either the fluctuations of $m$ and $h$ (see Fig.~\ref{flch1.71}).
\begin{figure}[ht]
\includegraphics[scale=0.5,angle=0]{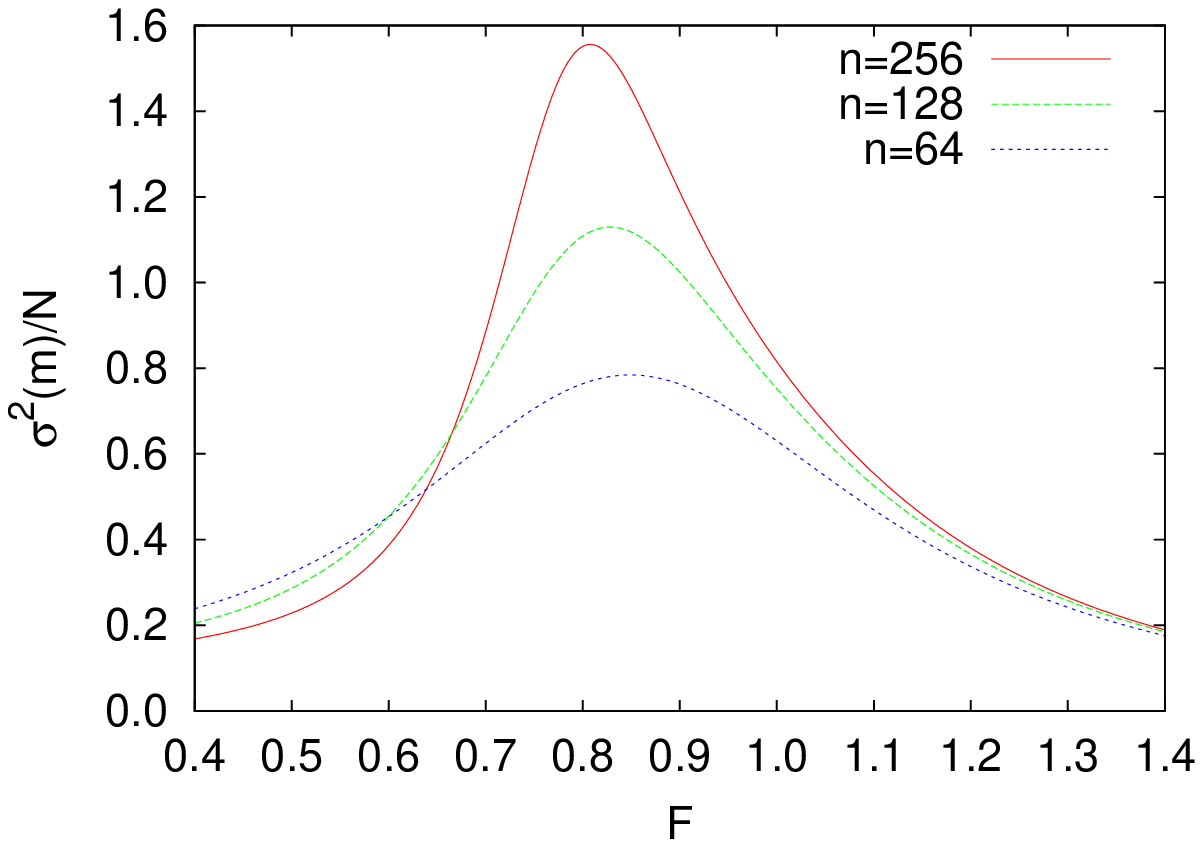}
\includegraphics[scale=0.5,angle=0]{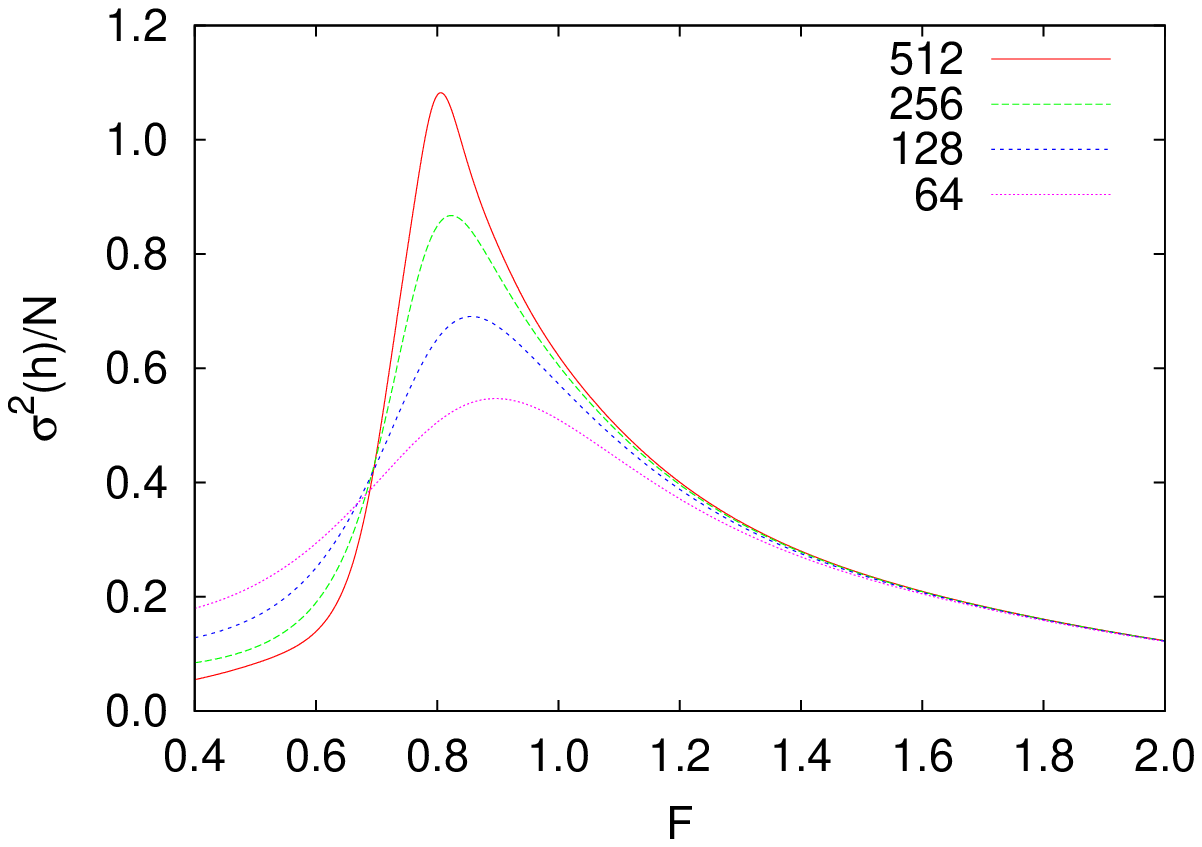}
\caption{\label{flcm0.7}  Plots of the fluctuations against force $F$
  for the low temperature $T=0.7$
}
\end{figure}

\begin{figure}[ht]
\includegraphics[scale=0.7,angle=0]{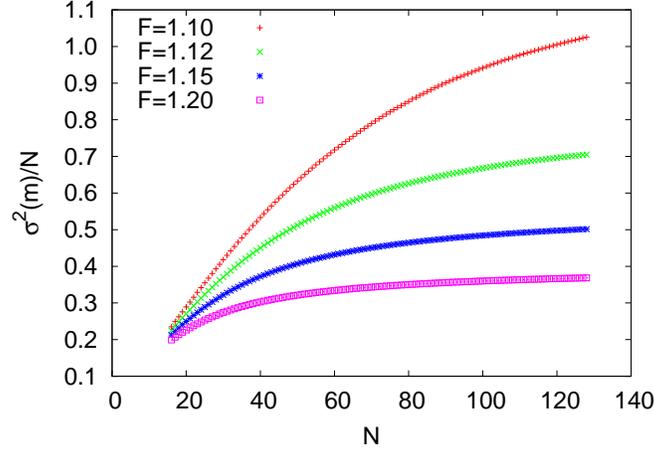} 
\caption{\label{max_flc}
The maximum of the fluctuations in the number of contacts $m$ for 
lengths from $N=32$ up to $N=128$ for four different forces in the
region where the original series data analysis detected a phase transition.}
\end{figure}

Now the question may be asked about the nature of the peaks in the fluctuations 
seen at low temperatures at forces just above $F=1$. In Fig.~\ref{max_flc} we
plot the maximum in the fluctuations at fixed force for various values
between $F=1.1$ and $F=1.2$. We note that while these peaks do
exist they are not indicative of any divergences. Of course there may
still be a weak phase transition. We now turn our attention to any
possible difference in the conformations of the polymer in the regions 
labeled ``stretched'' and ``fully stretched''.

\begin{figure}[ht!]
\includegraphics[scale=1.0,angle=0]{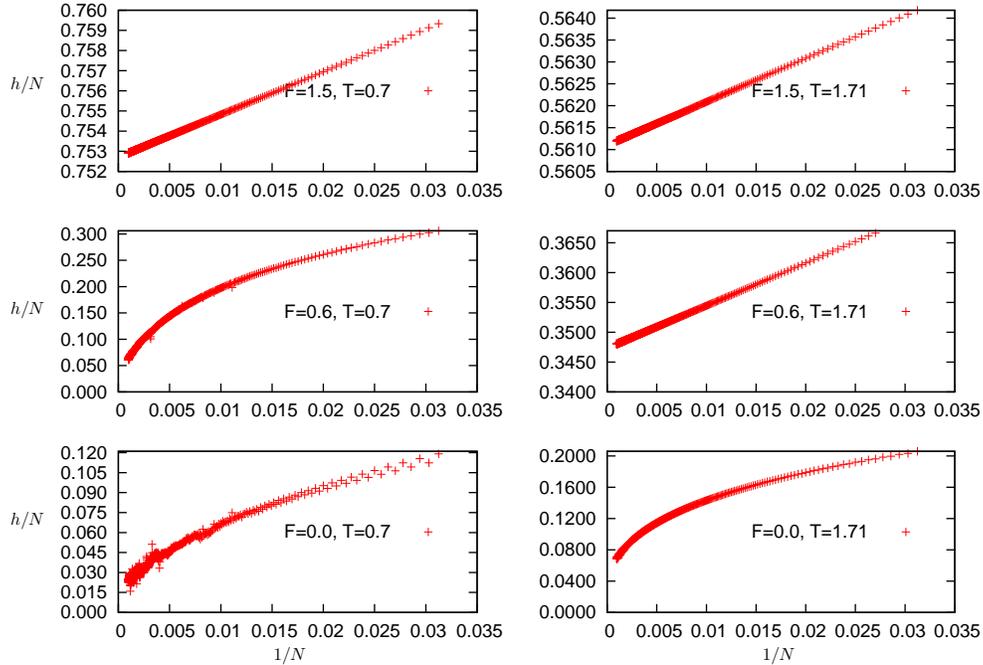}\\
\caption{\label{hN}
The height of the last monomer at six different points depicted in Fig.~\ref{pd_conj}. 
We see that in all cases the height of the last monomers does not converge to $1$, 
which we would expect for a fully stretched phase for figures at $F=1.5 ,T=0.7$ 
and $T=1.71$ and for $F=0.6,T=1.71$. 
}
\end{figure}

\begin{figure}[ht!]
\includegraphics[scale=0.7,angle=0]{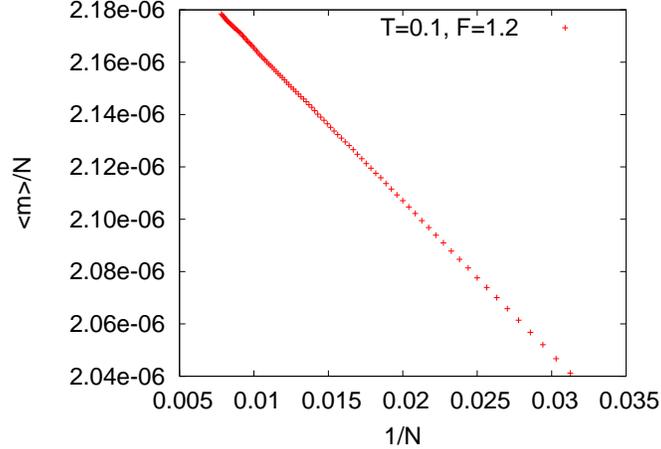} 
\caption{\label{scal_m} The behavior of the average number of bulk contacts 
$\left<m\right>$ for length $N=32$ up to $N=128$ for the point $T=0.1$, $F=1.2$ 
as a function of $1/N$. We observe an increase in the average number of contacts 
$\left<m\right>$ with $N$ even for very low temperatures.}
\end{figure}

To test the hypothesis on which the revised conjectured phase diagram
(Fig.~\ref{pd_inter}) rests we consider the scaling of the average height
of the last monomer. In Fig.~\ref{hN} we plot the height of the last 
monomer at six different points for temperatures $T=0.7$ and $T=1.71$.
We observe that at the three points
$(T,F)=(0.7,1.5), (1.71,1.5) \mbox{ and } (1.71,0.6)$ the average
height converges to a non-zero, and importantly, non-unity value. Also,
at the remaining three points, while there are clear non-linear corrections
to scaling, the average converges to zero. In other words no
indication of a fully stretched phase can be found. 
A further test of the hypotheses leading to the revised conjectured phase
diagram (Fig.~\ref{pd_inter}) can be carried out. We assumed that for very low
temperatures and large finite forces the average number of contacts per step
goes to zero. To test this we have plotted $\left<m\right>$ against $1/N$ for
$F=1.2$ with $T=0.1$ in Fig.~\ref{scal_m}. While small (of the order
of $10^{-6}$) this quantity is strictly increasing with length and
clearly converges to a (small) non-zero value.

\begin{figure}[ht]
\includegraphics[scale=0.7,angle=0]{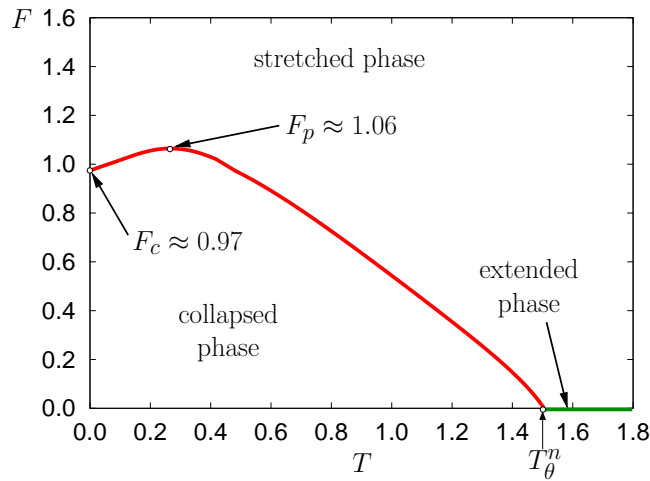}
\caption{\label{pd} 
  The finite-size phase diagram of a self-interacting self-avoiding
  walk under tension in two dimensions for length $N=128$.  We have estimated
  the position of the maximum force $F_p$, and the force $F_c$ for $T=0.0$. The
  $T^N_{\theta}\approx 1.47$. One would expect that $F_c=1.0$ in the
  thermodynamic limit.}
\end{figure}

We therefore conclude that the upper phase boundary proposed in 
\cite{jensen2007, jensen2008} does not exist in the thermodynamic limit
and the revised phase diagram in the thermodynamic limit is shown 
in Fig.~\ref{pd}. 

\begin{figure}[ht]
\includegraphics[scale=0.7,angle=0]{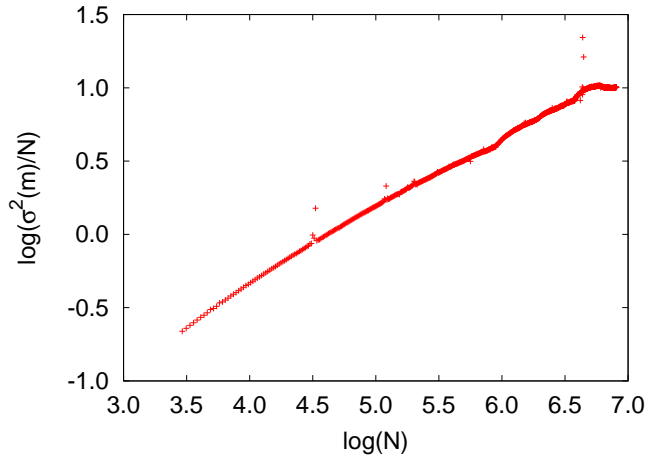}
\caption{\label{flcmT0.7} 
  The logarithm of the fluctuation in the number of contacts per monomer
  $m/N$ against $\log(N)$ at $T=0.074$ for the force $F$ at which the
  fluctuations are maximal. From this curve we can estimate the
  exponents $\alpha=0.62(10)$ and $\phi=0.72(6)$, which we note are not
those of the two-dimensional $\theta$-transition.}
\end{figure}

Finally we attempt to measure the exponents associated with the
collapse to stretched phase transition. This seems to be a second
order phase transition with divergent specific heat. In Fig.~\ref{flcmT0.7}
we plot the logarithm of the fluctuations in the number of contacts  
per monomer $m/N$ against $\log(N)$. The data in this plot is obtained 
at $T=0.074$ and at the force $F$ for which the fluctuations are maximal. 
From the date we obtain estimates of the specific heat exponent  
$\alpha=0.62(10)$  and the crossover exponent $\phi=0.72(6)$. The divergence 
of the finite size specific heat is expected to be controlled by the exponent
$\alpha\phi$ and the two exponents are expected to by related via 
the scaling relation $2-\alpha=1/\phi$.

\section{Summary  \label{sec:Summary}}

\begin{figure}[ht!]
\includegraphics[scale=0.6,angle=0]{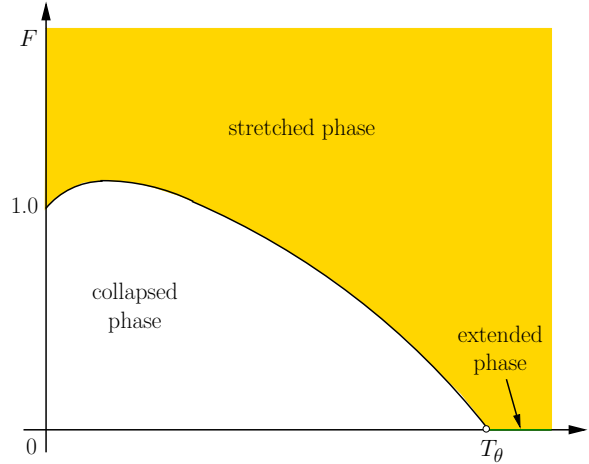}
\caption{\label{pd_schema_correct} 
The final conjectured phase diagram for a two-dimensional self-interacting 
polymer under a stretching force arising from our analysis in this work.}
\end{figure}

In summary we have shown that for a model of self-interacting polymers
pulled away from a surface in two dimension there are only two 
different phases for non-zero forces in the thermodynamic (infinite length) limit. 
We therefore conjecture a generic phase diagram as in Fig.~\ref{pd_schema_correct}.
One of the phases is the collapsed phase,
which is driven by the temperature at small forces. The other is a
single stretched phase which occurs whenever the force is applied for
temperatures higher than the $\theta$-temperature, and for large enough forces
for small temperatures. Importantly, the polymer is only in a fully
stretched state at zero temperature for forces $F\geq F_c=1$ or when
the applied force is infinite.

\section*{Acknowledgments}

The authors would like to thank Thomas Prellberg  for
helpful discussion.
Financial support from the Australian Research Council and the Centre
of Excellence for Mathematics and Statistics of Complex Systems is
gratefully acknowledged by the authors.  The exact enumerations were
performed on the computational resources of the Australian Partnership
for Advanced Computing (APAC), while the simulations were
performed on the computational resources of the Victorian Partnership
for Advanced Computing (VPAC). One of us (SK) would like to thank
the Department of Science and Technology and University Grants Commission, 
India for financial support.


\end{document}